\documentclass[10pt,aps,pre,twocolumn,superscriptaddress,floatfix,longbibliography]{revtex4-1}

\usepackage{graphicx}
\usepackage{amsmath}
\usepackage{amssymb}
\usepackage{hyperref}
\begin{document}

\title[Aging in two dimensional suspensions of rods as driven by Brownian diffusion]{Aging in two dimensional suspensions of rods as driven by Brownian diffusion}

\author{Nikolai I. Lebovka}
\email[Corresponding author: ]{lebovka@gmail.com}
\affiliation{Department of Physical Chemistry of Disperse Minerals, F. D. Ovcharenko Institute of Biocolloidal Chemistry, NAS of Ukraine, Kyiv, Ukraine, 03142}
\affiliation{Department of Physics, Taras Shevchenko Kyiv National University, Kyiv, Ukraine, 01033}
\author{Nikolai V. Vygornitskii}
\affiliation{Department of Physical Chemistry of Disperse Minerals, F. D. Ovcharenko Institute of Biocolloidal Chemistry, NAS of Ukraine, Kyiv, Ukraine, 03142}
\author{Yuri Yu. Tarasevich}
\email[Corresponding author: ]{tarasevich@asu.edu.ru}
\affiliation{Laboratory of Mathematical Modeling, Astrakhan State University, Astrakhan, Russia, 414056}
\date{\today}

\begin{abstract}
Aging in a two dimensional suspension containing rods was studied by means of kinetic Monte Carlo simulations. An off-lattice model with continuous positional and orientational degrees of freedom was considered. The initial state was produced using a random sequential adsorption model. During the aging, the rods underwent translational and rotational Brownian motions. The simulations were run at different values of number density (the number of rods per unit area), $\rho$, and of the initial orientation order parameter, $S_i$. The rods were assumed to have core---shell structures. The evolutions of both the connectivity and the order parameter in the course of the aging were examined.
\end{abstract}

\maketitle

\section{Introduction\label{sec:intro}}

Stability in colloidal suspensions of particles and the effects of aging in them are attracting a  great deal of attention~\cite{Cosgrove2010,Zheng2017,Tadros2017,Tadros2018}. In the initial state after preparation, the system is far away from thermodynamic equilibrium. During aging, different processes such as aggregation, the formation of arrested states, sol-gel  transitions and glass transitions can occur. Many intriguing phenomena in charged colloidal suspensions and suspensions of anisometric particles have been reported~\cite{Tanaka2005,Knaebel2000,Shahin2012}.

The aging effects in suspensions of impenetrable rodlike particles (rods) are of special interest.
For example, electrically conductive and transparent films prepared from sediments of carbon nanotubes are of particular interest for the production of electrodes for super-capacitors, thin film transistors, and fuel cells~\cite{Hu2010}.

For systems of elongated particles, different kinds of self-assembly and phase transition have been observed. For example, experimental studies have revealed the presence of jamming and orientational ordering during colloidal rod sedimentation~\cite{Mohraz2005}. Density-driven isotropic---nematic (IN) phase transition in three-dimensional (3D) homogeneous system of rods with infinite aspect ratio (length-to-diameter ratio $a=\infty$) was predicted theoretically in the 1940s~\cite{Onsager1949}. The theory predicted coexisting isotropic and nematic phases (for number densities of particles between $\rho_i\approx 3.29$ and $\rho_n\approx 4.19$), and transition to the nematic phase with strong ordering when $\rho\geq \rho_n$. Spontaneous ordering in concentrated rod suspensions has been experimentally confirmed in many studies~\cite{Adams1998,Maeda2003,Alargova2004}.

In two-dimensions (2D), the Onsager theory predicted a continuous IN transition at a critical density of $\rho_i=3\pi/2\approx 4.7$~\cite{Kayser1978}. Monte Carlo (MC) simulations have revealed the IN transition as the rod density, $\rho$, increases~\cite{Frenkel1985,Bates2000}. The ordered phase became absolutely unstable with respect to disclination unbinding at $\rho<\rho_i$, whereas at $\rho\geq\rho_n \approx 7.25$ a transition to a quasi-nematic phase occurred. The quasi-nematic phase demonstrated algebraic order (quasi long-range order) and the occurrence of a disclination-unbinding transition of the Kosterlitz---Thouless (KT) type has been suggested. MC simulations of continuous 2D systems of rods have revealed a KT type 2D nematic phase for relatively long rods with high aspect ratios of $a\gtrsim 7$ at high density~\cite{Bates2000}. 
The equilibrium steady state of hard rods in two dimensions has been studied using deposition-evaporation kinetics~\cite{Khandkar2005}. The simulation has confirmed that the critical number density is $\rho \approx 7$. The results of grand canonical simulation and accurate scaling analysis of the IN transition of hard needles in two spatial dimensions have confirmed that the transition is of the KT type~\cite{Vink2009}. The critical number density was found to be $\rho \approx 6.98$.

For non-equilibrium 2D systems of rods obtained using a random sequential adsorption (RSA) model, further self-assembly is possible. This self-assembly can be the result of a deposition-evaporation processes or of Brownian motion of the particles. Several problems related to such types of self-assembly of rods have previously been discussed~\cite{Ghosh2007,Lopez2010,Loncarevic2010,Matoz-Fernandez2012,Kundu2013,Kundu2013a,Lebovka2017}.

This paper analyzes the self-assembly and connectivity in 2D colloidal suspensions of rods during aging as driven by different aspects of Brownian motion. The initial state in our model was produced using an RSA model with anisotropic orientations of the rods. For the simulation of aging, the MC method was applied. During aging, the rods were allowed to undergo both translational and rotational motions. Note that in the limit of long times, MC produces trajectories equivalent to the trajectories of a system under Brownian dynamics. Generally, the standard MC approach is faster than using Brownian dynamics~\cite{Saintillan2005}. Proper rescaling of the MC time has allowed direct comparison between the simulations using MC and Brownian dynamics~\cite{Sanz2010,Romano2011,Patti2012,Cuetos2015}. Recently such approaches have been applied to the study of rodlike particles in the isotropic phase~\cite{Corbett2018}. The kinetics of the changes in the structure and connectivity of rods with core---shell structures have been analyzed.

The rest of the paper is constructed as follows. In Sec.~\ref{sec:methods}, the technical details of the simulations are described, all necessary quantities are defined, and some test results are given. Section~\ref{sec:results} presents our principal findings. Section~\ref{sec:conclusion} summarizes the main results.

\section{Computational model\label{sec:methods}}

The initial state  of the system under consideration was produced using an RSA model~\cite{Evans1993}. Rods of length of $l$ and thickness $d$ with large aspect ratios, ($a=l/d \gg 1$) were randomly and sequentially deposited onto a plane with periodic boundary conditions (i.e., onto a torus) until they reached the desired initial number density (concentration) $\rho_i$. Their overlapping with previously placed particles was forbidden. The dimension of the system under consideration was $L$ along both the horizontal direction $x$ and the vertical direction $y$. In the present work, all calculations were performed using $L=H=32 l$.

The MC procedure was used for simulation of the Brownian diffusion of the rods. The rotational and translational motions were taken into account. The rotational diffusion coefficient was calculated as $D_r=k_BT/f_r$. For the translational diffusion the motions both along and perpendicular to the direction of the long axis of the rods with diffusion coefficients of $D_\parallel=k_BT/f_\parallel$ and $D_\perp=k_BT/f_\perp$ were taken into account. Here $f_r$, $f_\parallel$, and $f_\perp$  are the Stokes' friction coefficients for the rotational and translational motions.

For long rods, $a\gg 1$,  the Stokes' friction coefficients can be evaluated, approximately, as~\cite{Loewen1994}
\begin{subequations}\label{eq:f}
\begin{align}
f_r &= \frac{\pi\eta l^3}{3(\ln a+\gamma_r)},\\
f_\parallel &= \frac{2\pi\eta l}{\ln a+\gamma_\parallel},\\
f_\perp &= \frac{4\pi\eta l}{\ln a+\gamma_\perp},
\end{align}
\end{subequations}
where $\eta$ is the viscosity of the surrounding liquid, and
$\gamma_r\approx -0.662$,
$\gamma_\parallel \approx -0.207$, and
$\gamma_\perp \approx 0.839$ are the end correction coefficients.

The amplitudes of the Brownian motions $\Delta \theta$,  $\Delta r_\parallel$, and $\Delta r_\perp$  are inversely proportional to the square root of the corresponding Stokes' friction coefficients $f_r$, $f_\parallel$, and $f_\perp$.
The amplitudes of the displacements were evaluated using the following equations
\begin{subequations}\label{eq:rt}
\begin{align}
\Delta r_\parallel  &= \beta l,\\
\Delta r_\perp  &=  \Delta r_\parallel \sqrt{f_\parallel/f_\perp},\\
\Delta \theta  &= \Delta r_\parallel/l \sqrt{f_\parallel/f_r}.
\end{align}
\end{subequations}
Here, the value of $\beta$ was chosen to be small enough ($\beta=0.05$) in order to obtain satisfactory acceptance of the MC displacement~\cite{Landau2014}. For example, at $a=10^3$, we have (see Eqs.~\eqref{eq:f}--\eqref{eq:rt})  $\Delta r_\perp/\Delta r_\parallel= 0.76 $ and  $l \Delta \theta /\Delta r_\parallel= 2.37\beta $.

One MC time step ($\Delta t_{MC}=1$) corresponds to two displacements and one rotation  attempted for all the rods in the system.  This time increment corresponds to the Brownian dynamics time increment that can be evaluated as~\cite{Patti2012}:
\begin{equation*}
\Delta t_{B}=\frac{\mathcal{A}_i}{3} \Delta t_{MC},
\end{equation*}
where $ \mathcal{A}_i $ is  the  acceptance coefficient for the $i$-th MC step.

The total Brownian dynamics time was evaluated as the sum
\begin{equation}
 t_{B}=\frac{ \Delta t_{MC} } {3}  \sum_{i=1}^{t_{MC}}\mathcal{A}_i,
\end{equation}
where $ t_{MC} $ is  the  MC time. Time counting was started from the value of $t_{MC}=0$, being the initial moment, and the total duration of the simulation was typically $\approx 10^6$ MC time units.

The orientation of the rods was characterized using the mean order parameter. This quantity was calculated as
\begin{equation}\label{eq:S}
S=\frac 1 N\sum\limits_{i=1}^{N} (2\cos^2 \theta_i -1).
\end{equation}
Here,  $\theta_i$ is the angle between the axis of the $i$-th rod and the horizontal axis $x$, while $N$ is the total number of rods.

For ideally oriented rods along the horizontal axis $x$, $S=1$, and for isotropic initial orientation of the rods, $S=0$. In the course of Brownian motion, the axis of preferred orientation was determined and the order parameter $S$ was calculated by taking account of the orientation of the rods relative to this axis.

At the initial moment, the rods were aligned with respect to a selected direction, $x$. Their axes were uniformly distributed within some  interval such that, $-\theta_m \leq \theta\leq \theta_m$, where $\theta_m\leq \pi/2$. The isotropic case is given by $\theta_m=\pi/2$, and the smaller the value of $\theta_m$ the higher is the degree of orientation. This model was equivalent to that considered in~\cite{Balberg1983,Balberg1983a,Balberg1984}. In this case, the initial order parameter can be evaluated as
\begin{equation}\label{eq:stheta1}
   S_i  = \frac{2}{\theta_m}\int_{0}^{\theta_m} \cos^2 \theta \, d\theta -1 = \frac{\sin 2\theta_m} {2\theta_m}.
\end{equation}

\begin{figure}[!htb]
  \centering
 \includegraphics[width=0.75\columnwidth]{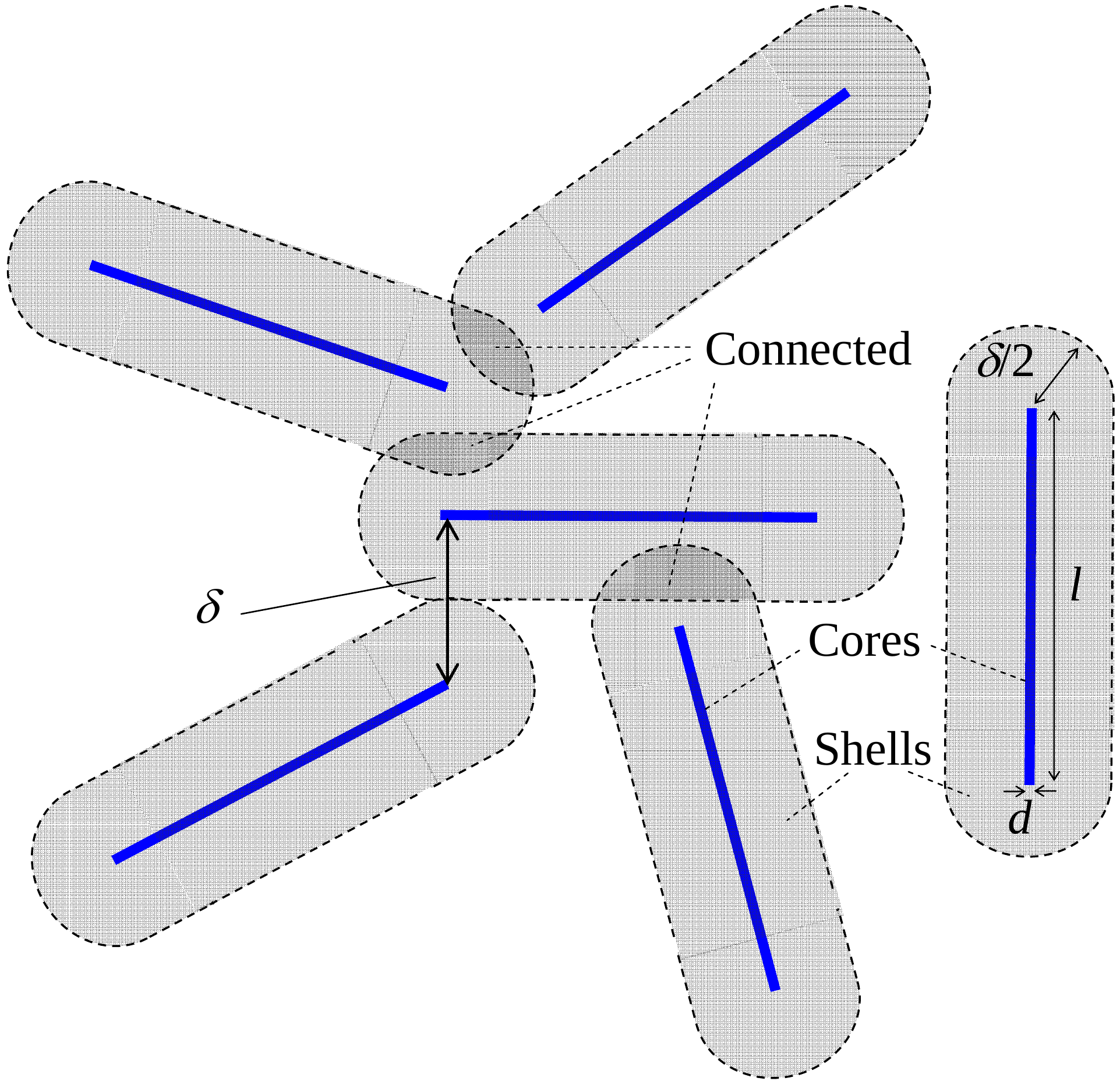}
  \caption{Sketch to the definition of the connectivity of rods with a core---shell structure. The particles are assumed to be connected when the distance between them does not exceed the value of $\delta_s$. \label{fig:Connectf01}}
\end{figure}

To characterize the 2D film, the connectivities of particles along the $x$ and $y$ directions were estimated. In these calculations, the core---shell structure of the particles was assumed. The minimum (critical) values of the outer shell thicknesses $\delta_x$ and $\delta_y$ required for the formation of spanning clusters along the $x$ and $y$ directions, respectively, were evaluated using the Hoshen---Kopelman algorithm \cite{Marck1997}. The particles were assumed to be covered by outer shells with a thickness of $\delta_s/2$ and to be connected when the distance between them does not exceed the value of $\delta_s$ (Fig.~\ref{fig:Connectf01}). The analysis was performed using the near-neighbor particles from the list, while, for distances smaller than the value of $2l$, takin into account the periodic boundary conditions. The shortest distance between rods was evaluated using the fast algorithm proposed in  \cite{Vega1994}. The value $\delta_{xy}$ corresponds to the minimum thickness for percolation along the $x$ or $y$ directions. The connectivity anisotropy was defined as
\begin{equation}
 \Delta=\delta_{y}/\delta_{x}. \label{Eq:Anisotropy}
\end{equation}

\begin{figure*}[!hbt]
  \centering
  \includegraphics[width=0.95\linewidth]{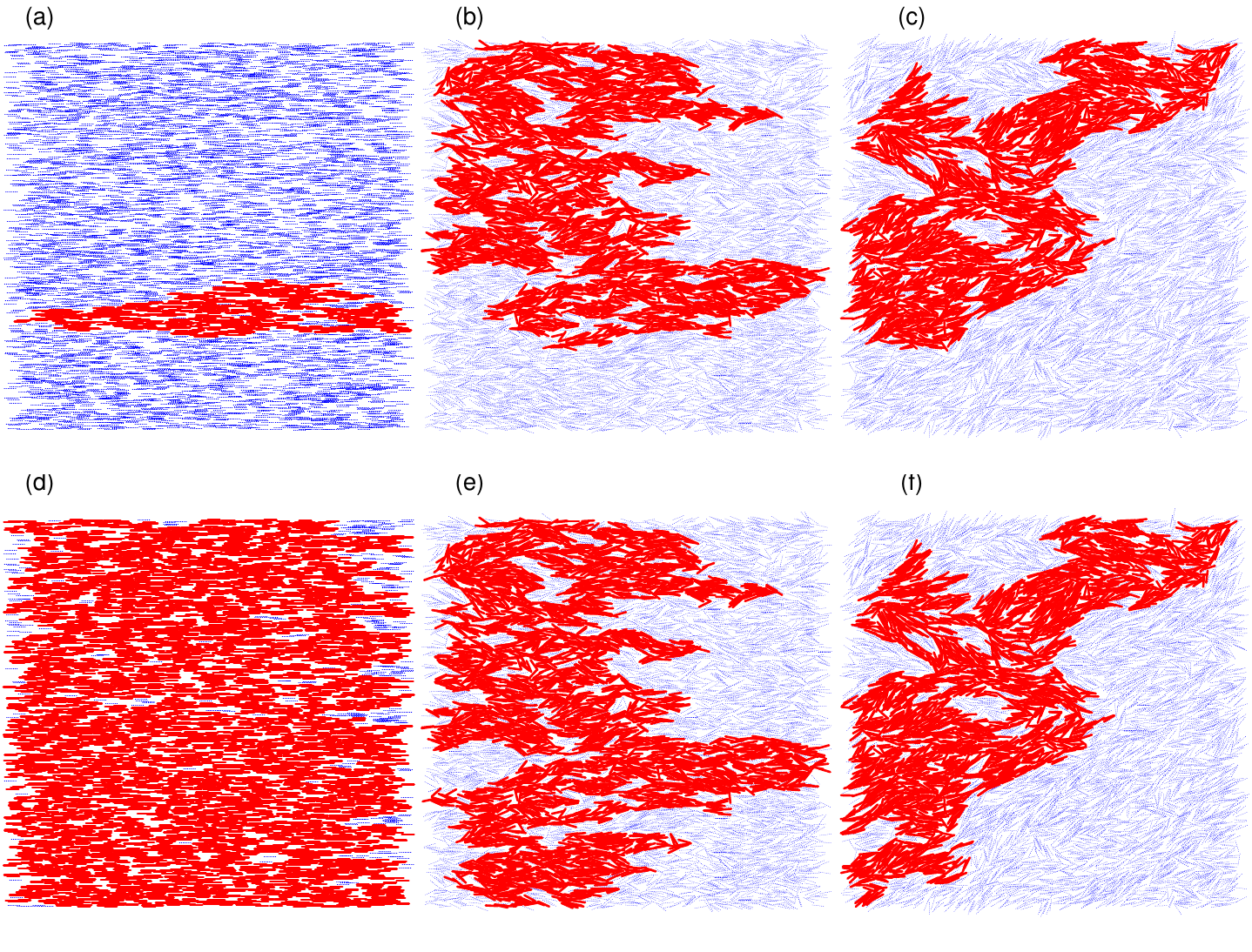}
  \caption{Example of the evolution of patterns at different times, $t_B=0$ (a,d), $t_B=250$ (b,e), and $t_B=3\times 10^5$ (c,f), at an initial order parameter of $S_i=1$, for a number density of rods of $\rho=7$. The marked particles correspond to the connected clusters along the $x$ direction (upper row) and $y$ direction (bottom row). \label{fig:Patternsf02}}
\end{figure*}

The values of $\delta_x$ and $\delta_y$ were evaluated using the bisection method with errors that did not exceed the value of $10^{-4}$. The value $\delta$ corresponds to the minimum thickness for percolation along the $x$ or $y$ directions. Figure~\ref{fig:Patternsf02} presents examples of evolution of the connectivity patterns at different Brownian times at an initial order parameter of $S_i=1$ and a number density of rods of $\rho=7$\footnote{See Supplemental Material at [URL will be inserted by publisher] for an animation of the temporal evolution of patterns  at an initial order parameter of $S_i=1$, for a number density of rods of $\rho=7$.}. The marked particles correspond to the connected clusters along the $x$ direction (upper row) and the $y$ direction (bottom row).

For given value of $\delta$ ($\delta_{y}$ or $\delta_{x}$) the effective aspect ratio of a particle can be evaluated as $a=(l+2\delta)/(2\delta)=1+l/(2\delta)$.
The excluded area $S_{ex}$ was evaluated as~\cite{Balberg1984a}
\begin{equation}\label{eq:Vex2D}
S_{ex}=l^2\left[\langle \sin\psi \rangle+4/a+\pi/a^2\right],
\end{equation}
where $\psi$ is the angle between the two particles, and $\langle \cdot\rangle$ corresponds to the number-averaged value.

For random orientations of particles ($S=0$),  $\langle \sin\psi \rangle=2/\pi$, while, for ideally oriented particles ($S=1$), $\langle \sin\psi \rangle=0$, and in the general case the $\langle \sin\psi \rangle (S)$ dependence may be well approximated with a determination coefficient of $R^2=0.99998$ by the equation
\begin{equation}\label{eq:Theta2D}
\langle \sin\psi \rangle=\frac{2}{\pi}\sqrt{1+aS+BS^2+cS^3},
\end{equation}
where $a=-0.153 \pm 0.001$, $b=-1.084 \pm 	0.002$, and $c=0.230 \pm 0.001$.

For particles with overlapping shells, the total fraction of the space covered by the particles with shells (the filling fraction) was evaluated as~\cite{Mertens2012}
\begin{equation}\label{eq:phi}
\varphi=1-\exp(-\rho S_s),
\end{equation}
where $S_s=l\delta+\pi\delta^2/4$ is the total area of each individual particle covered with a shell.

For each given value of $\rho$ or $S$, the computer  experiments were repeated up to $100$ times.  The error bars in the figures correspond to the standard deviation of the mean. When not shown explicitly, they are of the order of the marker size.

\section{Results and Discussion\label{sec:results}}

\begin{figure}[!htb]
  \centering
 \includegraphics[width=0.95\columnwidth]{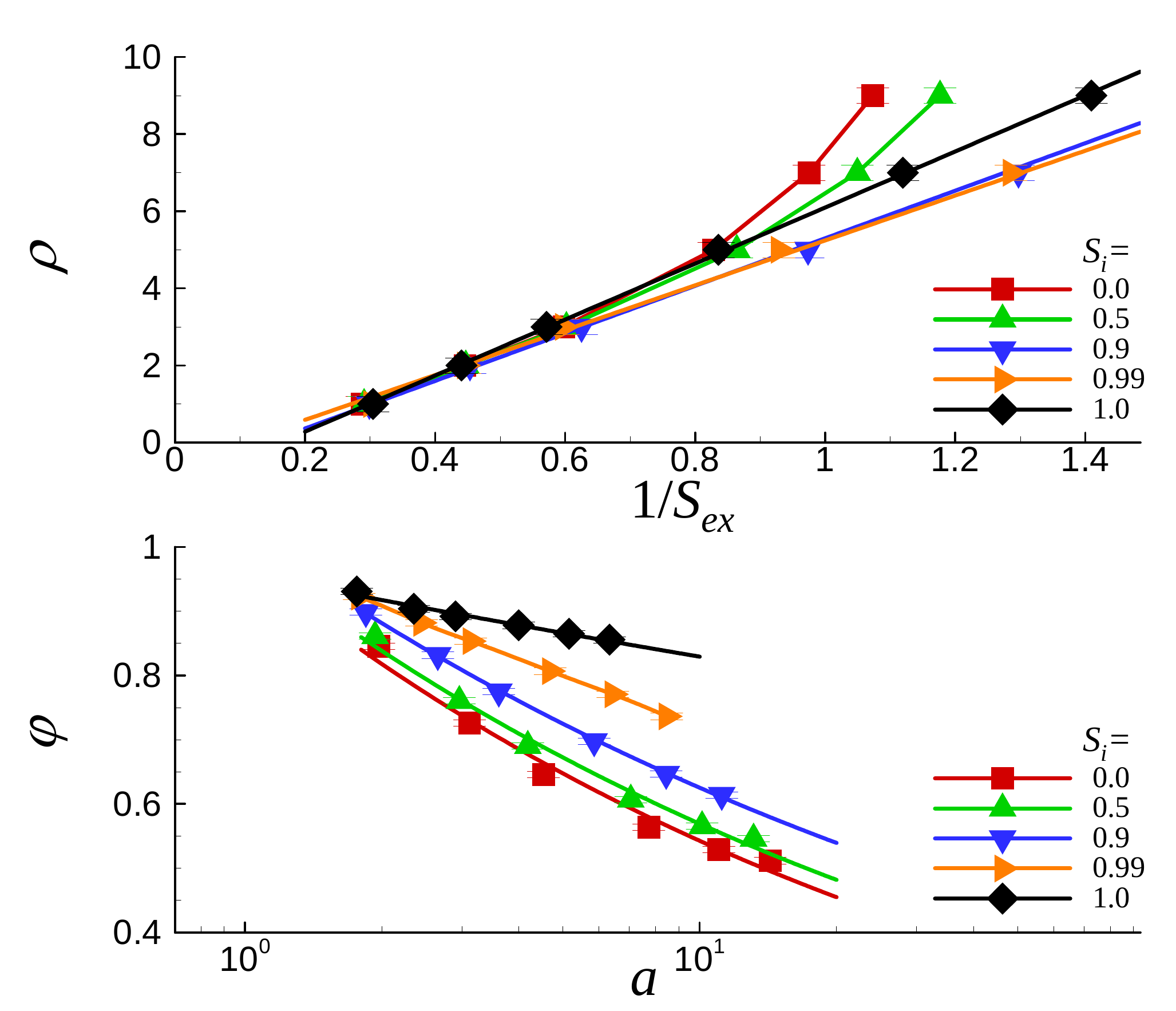}
  \caption{Number density, $\rho$, versus inverse excluded area, $1/S_{ex}$, (a) and  the filling fraction, $\varphi$ versus the effective aspect ratio, $a$, (b) for different values of the order parameter, $S_i$, at the initial time, $t_B=0$. In these calculations, we used the minimum thickness of the outer shell, $\delta_{x}$.\label{fig:Deltaf03}}
\end{figure}

Figure~\ref{fig:Deltaf03} presents the number density, $\rho$, versus the inverse of the excluded surface area, $1/S_{ex}$, (a) and  the filling fraction, $\varphi$ versus the effective aspect ratio, $a$, (b) for different values of the order parameter, $S_i$, at the initial time, $t_B=0$. In these calculations, we used the minimum thickness of the outer shell, $\delta_{x}$. For highly ordered systems with $S_i\geq 0.9$, the linear proportionality between $\rho$ and $1/S_{ex}$ can be observed (Fig.~\ref{fig:Deltaf03}a) is in full correspondence with conjecture of theory~\cite{Balberg1984a}. However, for disordered systems ($S<0.9$) a significant deviation from this rule was observed for high number densities above $\rho\approx 5$. The filling fraction, $\varphi$ continuously decreased with  the effective aspect ratio, $a$, (Fig.~\ref{fig:Deltaf03}b) and this effect was more pronounce for disordered systems. A similar impact of the aspect ratio on the critical filling fraction  at percolation points has also been observed for randomly oriented ellipses~\cite{Xia1988} and rectangles~\cite{Li2013}.

\begin{figure}[!htb]
  \centering
 \includegraphics[width=0.95\columnwidth]{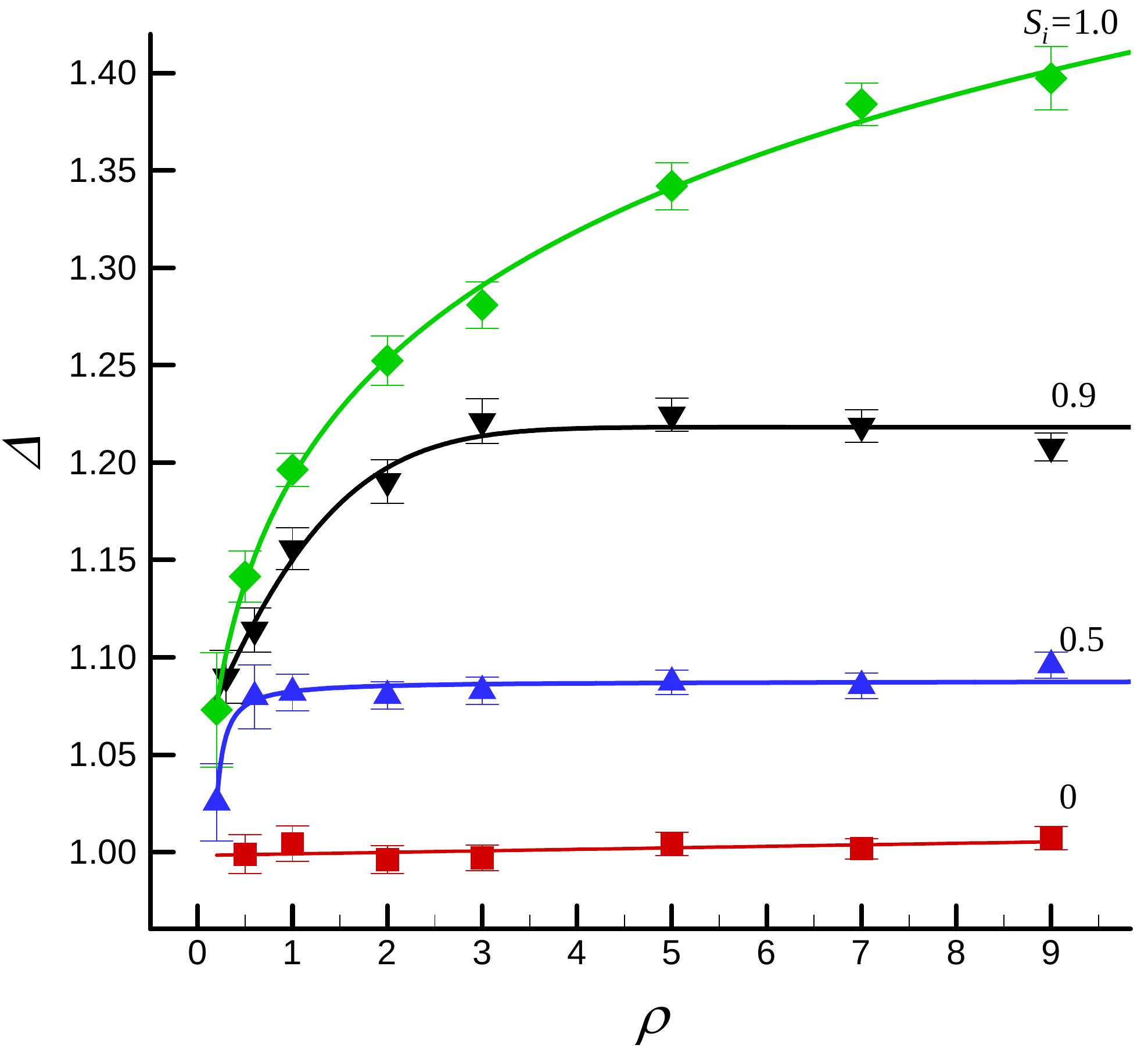}
  \caption{Connectivity anisotropy, $\Delta=\delta_y/\delta_x$,	 versus the number density of rods $\rho$ for different values of the order parameter, $S_i$, at the initial time, $t_B=0$. \label{fig:Anisotropy04}}
\end{figure}

For any values of $S$, the values of $\delta_y$ exceeded the values of $\delta_x$. Figure~\ref{fig:Anisotropy04} shows the connectivity anisotropy, $\Delta$, versus the number density of rods $\rho$ for different values of the order parameter, $S$, at the initial time, $t_B=0$. For isotropic system at $S_i=0$, we have $\Delta=1$. For anisotropic systems at $S_i>0$, the values of $\Delta$ increased initially with $\rho$ but then stabilized at large values of $\rho$.

\begin{figure}[!htb]
  \centering
 \includegraphics[width=0.95\columnwidth]{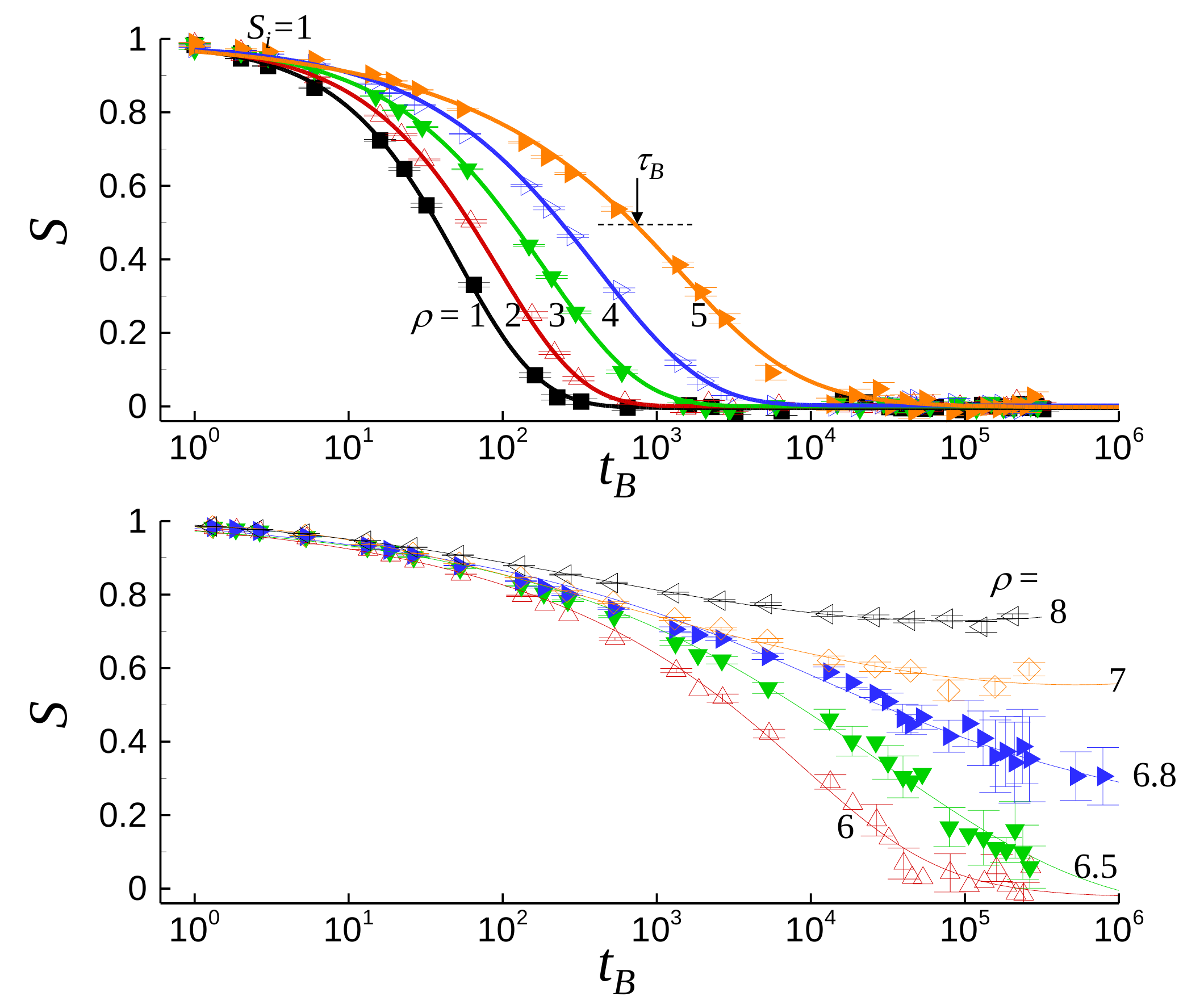}
  \caption{Order parameter, $S$, versus Brownian time, $t_B$, at  different number densities of rods, $\rho$. At the initial time, $t_B=0$ the order parameter was $S_i=1$. \label{fig:S1_05}}
\end{figure}

Figure~\ref{fig:S1_05} shows the order parameter, $S$, versus Brownian time, $t_B$, at  different number densities of rods, $\rho$. These data were obtained starting from ideally aligned systems at the initial time, $t_B=0$, i.e., with $S_i=1$. For relatively small values of $\rho$ ($\rho<6.7$), the ideal order was destroyed after some time, while at ($\rho\geq 6.7$) the systems remained  stabilized in quasi-nematic states with a finite order parameter $S_\infty$.  We can defined the characteristic Brownian time $\tau_B$ as the time required to reach the level of $(1-S_\infty)/2$.

\begin{figure}[!htb]
  \centering
 \includegraphics[width=0.95\columnwidth]{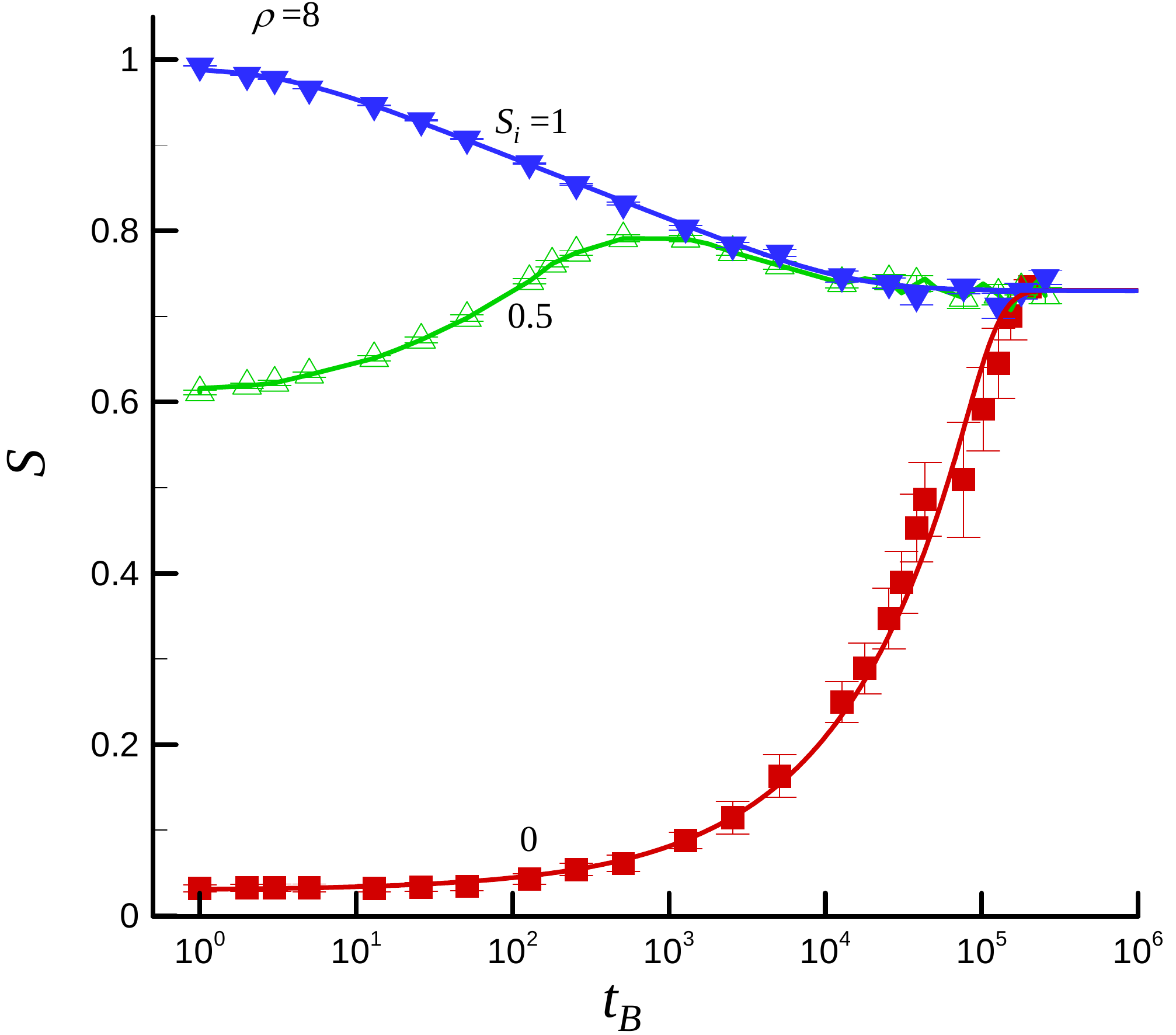}
  \caption{Examples of the evolution over time of the order parameter, $S$, at  different values of the initial order parameters, $S_i$, for the number density of rods, $\rho=8$. \label{fig:Ro8_06}}
\end{figure}

Figure~\ref{fig:Ro8_06} shows examples of the order parameters, $S$, versus the Brownian time, $t_B$, at  different values of the initial order parameter, $S_i $ and a fixed number density of rods, $\rho=8$. For this system, all $S(t_B)$ curves converged at the level of $S_\infty \approx 0.69$. For a system started from a completely disordered initial state ($S_i=0$), the order parameter $S$ was very low during the initial aging and started to grow only when $t_B>10^4$. Surprisingly, for the intermediate value of $S_i=0.5$, the $S(t_B)$ curve went through its maximum at $t_B \approx10^3$ at that point approaching the $S(t_B)$ curve for $S_i=1$.

\begin{figure}[!htb]
  \centering
 \includegraphics[width=0.95\columnwidth]{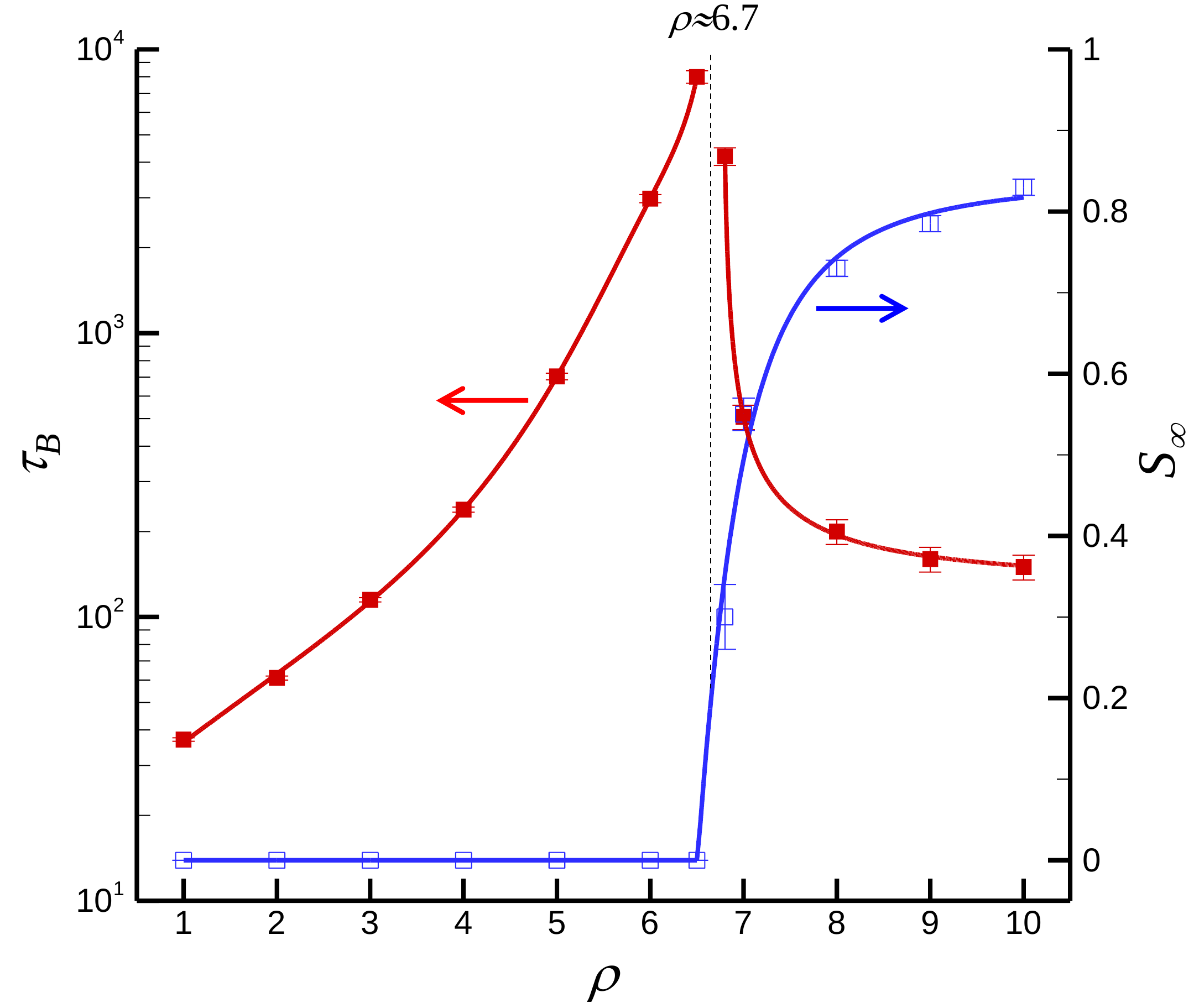}
  \caption{Characteristic Brownian time, $\tau_B$, and saturation level of the order parameter, $S_\infty$, versus the number density of rods $\rho$. \label{fig:TauSoo_07}}
\end{figure}

Figure~\ref{fig:TauSoo_07} shows the characteristic Brownian time, $\tau_B$, and the saturation level of the order parameter, $S_\infty$, versus the number density of rods of $\rho$. The value of $\tau_B$ significantly increased with $\rho$ in the disordered phases at $\rho<6.9$ ($S_\infty=0$). However, $\tau_B$ decreased with $\rho$ in the quasi-nematic phases at $\rho>6.9$. The order parameter $S_\infty$ in the quasi-nematic phases continuously grew with $\rho$ (Fig.~\ref{fig:TauSoo_07})

\begin{figure}[!htb]
  \centering
 \includegraphics[width=0.95\columnwidth]{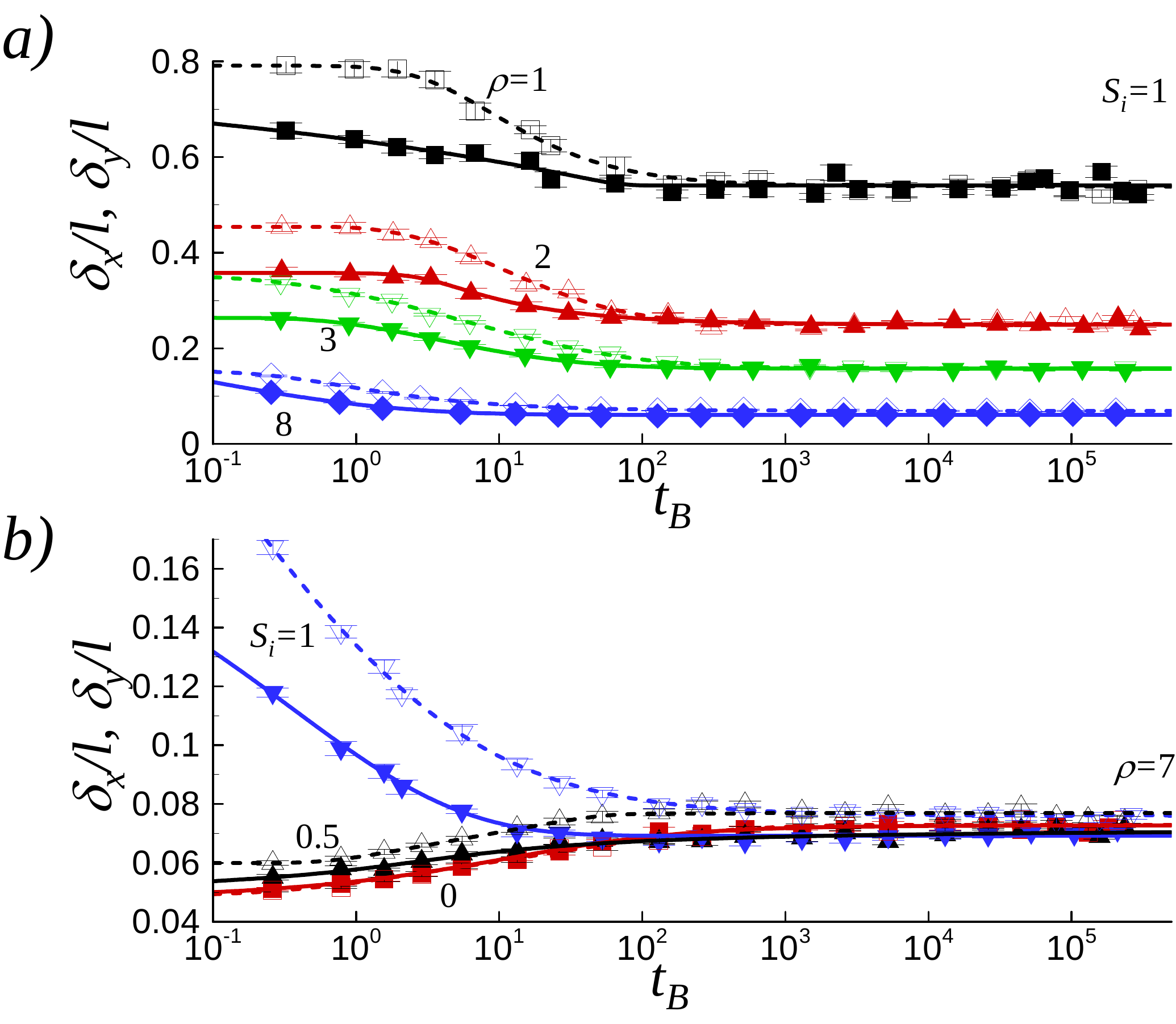}
  \caption{Changes in the connectivities $\delta_x/l$ and $\delta_y/l$ during aging for fixed value of the initial order parameter $S_i=1$ and different number densities of rods, $\rho$, (a), and for fixed value of $\rho=7$ and different values of $S_i$ (b). Solid lines, filled symbols and dashed lines, open symbols correspond to the $\delta_x$ and $\delta_y$ values respectively.  \label{fig:Deltaxy_08}}
\end{figure}

Figure~\ref{fig:Deltaxy_08} demonstrates the changes in the connectivities $\delta_x/l$ and $\delta_y/l$ during aging for the fixed value of the initial order parameter $S_i=1$ and different number densities of rods, $\rho$, (a), and for the fixed value of $\rho=7$ with different values of $S_i$ (b). For fixed values of $S_i=1$ (Fig.~\ref{fig:Deltaxy_08}a), the data evidenced that, at the initial time, the values of $\delta_y(t_B)$ exceeded the values of $\delta_x(t_B)$, so significant anisotropy in percolation could be observed. However, fast convergence of both the $\delta_x(t_B)$ and $\delta_y(t_B)$ dependencies during the initial aging time ($t_B\leq 10^2$) was observed. Similar behavior was observed for the fixed value of $\rho=7$ and different values of $S_i$ (Fig.~\ref{fig:Deltaxy_08}b).

\section{Conclusion\label{sec:conclusion}}
A continuous 2D model of Brownian motion-driven aging in suspensions of rods was studied by MC simulation. The initial state  was produced using an RSA model with anisotropic orientations of the rods. During aging, the rods undergo translational and rotational Brownian motions. At the initial time, $t_B=0$, for anisotropic systems at $S_i>0$ the connectivity anisotropy, $\Delta$,	increased with the particle number density, $\rho$. The aging behavior of perfectly aligned systems (with $S_i=1$ at the initial time, $t_B=0$) was dependent upon the concentration of the rods. At low values such that $\rho<\rho_n$ ($\rho_n\approx 6.7$), the ideal order was destroyed after some time, while at $\rho\geq \rho_n$ the systems were stabilized in quasi-nematic states with finite order parameter, $S_\infty$. The estimated value of $\rho_n\approx 6.7$ is slightly smaller than the previously estimated the critical density $\rho_n \approx 6.98-7.25$ for the transition to a quasi-nematic phase~\cite{Frenkel1985,Bates2000,Khandkar2005,Vink2009}. The characteristic Brownian time $\tau_B$ required to transition to the equilibrium state significantly increased with $\rho$ in the disordered phases and decreased with $\rho$ in the quasi-nematic phases. The order parameter $S_\infty$ in the quasi-nematic phases grew continuously with $\rho$. At the initial time ($t_B<100$) the values of $\delta_y(t_B)$ exceeded the values of $\delta_x(t_B)$, so significant anisotropy in percolation was observed.

\section*{Acknowledgments}
We acknowledge funding from the National Academy of Sciences of Ukraine, Projects No.0117U004046 and~43/19-H (N.I.L., N.V.V.), and the Ministry of Science and Higher Education of the Russian Federation, Project No.~3.959.2017/4.6 (Yu.Yu.T.).

\bibliography{Aging}

\end{document}